\documentstyle{article}

\author{Hao Chen\\
Department of Mathematics\\
Zhongshan University\\
Guangzhou,Guangdong 510275\\
People's Republic of China}
\title{\bf Necessary conditions for efficient simulation of Hamiltonians using local unitary operations}
\date{August,2001}

\begin{document}

\maketitle
\begin{abstract}
We give necessary conditions for the efficient simulation of both bipartite and multipartite Hamiltonians, which are 
independent of the eigenvalues of Hamiltonians and based on the algebraic-geometric invariants introduced in [1] and 
[2]. The results show that the problem of efficient simulation of Hamiltonians on arbitrary bipartite or multipartite 
quantum systems cannot be described by only using eigenvalues, which is quite different to the two-qubit case.

\end{abstract}

Historically the idea of simulating Hamiltonian time evolutions was the first motivation for quantum computation [3].
 Recently the ability of nonlocal Hamiltonians to simulate one another is a 
popular topic , which has applications in quantum control theory [4], quantum computation [5],[6],[7],[8] and the 
task of generating entanglement [9] [10]. The problem to parameterize the nonlocal properties of interaction 
Hamiltonians, so as to characterize the efficiency with which they can be used to simulate one another, 
is theoretically and experimentally important. There have been very active research on this problem ([11],
[12],[13],[14],[15],[16],[17]). For the general treatments of this topic , we refer to [11] as the main reference.\\

In [11] it was  shown that the efficiency with which Hamiltonian $H$ , together with local operations , simulates 
another 
Hamiltonian $H'$ can be used as a criterion to endow the set of Hamiltonians with a  partial order structure, that 
allows 
to compare the nonlocal capabilities of $H$ and $H'$. For two-qubit Hamiltonians, it was  shown that the problem 
of simulation of Hamiltonians can be reduced to the case of so-called normal forms of these Hamiltonians (Theorem 
in section V A of [11]). For these Hamiltonians in their normal forms , a beautiful necessary and sufficient 
condition about the simulating capability in terms of the majorization of eigenvalues of these Hamiltonians 
was given. This indicated that the partial order structure endowed on the two-qubit Hamiltonians is in close analogy 
to the partial ordering of bipartite pure states endowed by their capabilities to be converted by LOCC ([18]).\\

It is natural to consider  the simulation problem of Hamiltonians on arbitrary bipartite quantum systems. We can 
imagine that in higher dimensions, Hamiltonians have more nonlocal degrees of freedom than the two qubit case, and 
there is no result about "normal forms" of Hamiltonians on arbitrary bipartite quantum systems, this may make the 
problem more difficult and it seems hopeless to give a characterization based only on the majorization of some 
numerical quantities (as in Theorem of section F of [11]). In [19] it is proved that $m^2n^2-m^2-n^2+1$ nonlocal 
parameters are needed to describe the set of equivalent classes of bipartite mixed states on $H_A^m \otimes H_B^n$ 
under local unitary operations. A similar parameter counting argument as [19] shows that there must be at least 
$m^2n^2-2(m^2+n^2)+3$ nonlocal parameters for the equivalent classes of bipartite Hamiltonians on $H_A^m \otimes 
H_B^n$ under local unitary operations. When we consider the ability of nonlocal Hamiltonians to simulate one another 
with the help of local unitary operations, it is natural to imagine that these continuous invariants of Hamiltonians 
under local unitary operations may give constraints on these Hamiltonians $H$ and $H'$ if there is a simulation 
relation using local unitary operations between them ,and any such constraint must be expressed by these invariants. 
In our previous works [1] and [2] the algebraic-geometric invariants of bipartite mixed states (i.e. semi-positive 
self-adjoint operators) were introduced as their nonlocal invariants (i.e., these algebraic sets are kept invariant 
under local unitary operations). These algebraic-geometric invariants depends only on eigenvectors and are independent 
of eigenvalues of the semi-positive adjoint operators. We can think these algebraic-geometric invariants as nonlocal 
invariants of semi-positive bipartite Hamiltonians and ask if there exists any constraint on these invariants of two 
semi-positive Hamiltonians $H$ and $H'$ if $H$ can be simulated by $H'$ using local unitary operations. \\

In this paper, we show that the efficient simulation relation between two semi-positive bipartite Hamiltonians of the 
same rank implies the equalities of these algebraic-geometric invariants of them. This necessary condition is also 
extended to the efficient simulation of multipartite Hamiltonians. Since these algebraic-geometric invariants are 
independent of the eigenvalues and only measure the position of eigenvectors of the Hamiltonians. Thus our results 
strongly suggest that the eigenvectors play a more fundamental role in the efficient simulation problem of Hamiltoians 
on arbitrary bipartite or multipartite systems. This is quite different to the two qubit case studied in [11].\\

In this paper, we say, for two bipartite Hamiltonians $H$ and $H'$ on $H_A^m \otimes H_B^n$, $H'$ can be efficiently 
simulated by $H$ with local unitary operations, write as $H' \prec_{LU} H$, if $H'$ can be written as a convex 
combination of conjugates of $H$ by local unitary operations, 
$H'=p_1(U_1 \otimes V_1)H(U_1 \otimes V_1)^{\dagger}+...+p_s (U_s \otimes V_s) H(U_s \otimes V_s)^{\dagger}$, 
where $p_1,...,p_s$ are positive real numbers such that $p_1+...+p_s=1$ , $U_1,...,U_s$ and $V_1,...,V_s$ are 
unitary operations on $H_A^m$ and $H_B^n$ respectively. Here we use $\dagger$ for the adjoint. This is equivalent to 
the notion
"infinitesimal simulation" in [11] and "first order simulation in [16]. In [11] and [16] it is shown that "local 
terms" like $I \otimes K_B$ and $K_A \otimes I$ are irrelevant to the simulation problem upto the second order, thus 
they consider the simulation problem for Hamiltonians without local terms' effect. Our definition here is more 
restricted without neglecting the local terms.\\

If $H' \prec_{LU} H$, where $H$ is a semi-positive self-adjoint operator, it is clear that $H'$ has to be a 
semi-positive self-adjoint operator and $tr(H')=tr(H)$. Thus it is clear that $H' \prec_{LU} H$ is equivalent to 
$H'-\frac{tr(H')}{mn}I_{mn} \prec_{LU} H-\frac{tr(H)}{mn}I_{mn}$, so we do not restrict to the traceless Hamiltonians 
in this paper. \\
 
We have the following observation. First we recall the following result in [20].\\

{\bf Lemma 1.} {\em Let $T=\Sigma_i p_i |v_i \rangle \langle v_i|$, where $p_i$'s are positive real numbers, be a 
positive self-adjoint operator on a finite dimensional Hilbert space. Then the range $range(T)$ of $T$ is the linear 
span of vectors $v_i$'s.}\\

If $H=|v \rangle \langle v|$ and $H'=|v'\rangle \langle v'|$  where $|v \rangle $ and $|v'\rangle$ are pure states and 
$H'$ can be simulated by $H$ efficiently, ie., $H' \prec_{LU} H$, actually the Schmidt ranks of $|v \rangle $ and $|v' 
\rangle$ have to be the same. In fact, if there exist positive numbers $p_1,...,p_s$ and local unitary operations $U_1 
\otimes V_1,...,U_s \otimes V_s$, such that, $\Sigma_i p_i U_i \otimes V_i H (U_i\otimes V_i)^{\dagger}=H'$, it is 
clear that $U_i \otimes V_i H (U_i\otimes V_i)^{\dagger}=|(U_i \otimes V_i)v \rangle \langle (U_i \otimes V_i)v|$, and 
from Lemma 1, $|(U_i \otimes V_i)v \rangle$ is in the range of $H'$. Hence $|v' \rangle=|(U_i \otimes V_i)v \rangle$ 
and the Schmidt ranks of $|v \rangle$ and $|v'\rangle$ have to be the same.\\

For semi-positive bipartite Hamiltonians (equivalently ,bipartite mixed  states, i.e., semi-positive self-adjoint 
operators) on $H_A^m \otimes H_B^n$, algebraic sets $V_A^k(H)$ in $CP^{m-1}$ (respectively $V_B^k(H)$ in $CP^{n-1}$) 
are introduced in [1] as the {\em degenerating locus} of the measurement of them by separable pure states. For any 
given semi-positive self-adjoint operator (bipartite mixed states or semi-positive Hamiltonians) $\rho$ on $H_A^m 
\otimes H_B^n$ , we consider the expression $\langle \phi_1 \otimes \phi_2 |\rho|\phi_1 \otimes \phi_2\rangle$ for any 
pure states $\phi_1 \in H_A^m$ and $\phi_2 \in H_B^n$. For any fixed $\phi_1 \in P(H_A^m)$, where $P(H_A^m)$ is the 
projective space of all pure states in $H_A^m$,  $\langle \phi_1 \otimes \phi_2 |\rho|\phi_1 \otimes \phi_2 \rangle$ 
is a Hermitian bilinear form on $H_B^n$, denoted by $ \langle \phi_1|\rho|\phi_1 \rangle$ . We consider the {\em 
degenerating locus } of this bilinear form, ie., $V_A^k(\rho)=\{\phi_1 \in P(H_A^m): rank (\langle \phi_1|\rho|\phi_1 
\rangle) \leq k\}$ for $k=0,1,...,n-1$. We can use the coordinate form of this formalism. Let 
$\{|11\rangle,...,|1n\rangle,...,|m1\rangle,...,|mn\rangle\}$ be the standard orthogonal basis of $H_A^m \otimes 
H_B^n$ and $\rho$ be an arbitrary (semi)positive self-adjoint operator. We represent the matrix of $\rho$ in the basis 
$\{|11\rangle,...|1n\rangle,...,|m1\rangle,...,|mn\rangle\}$, and consider $\rho$ as a blocked matrix 
$\rho=(\rho_{ij})_{1 \leq i \leq m, 1 \leq j \leq m}$ with each block $\rho_{ij}$ a $n \times n$ matrix corresponding 
to the $|i1\rangle,...,|in\rangle$ rows and the $|j1\rangle,...,|jn\rangle$ columns. For any pure state 
$\phi_1=r_1|1\rangle+...+r_m|m\rangle \in P(H_A^m)$ the matrix of the Hermitian linear form 
$\langle\phi_1|\rho|\phi_1\rangle$ with the basis $|1\rangle,...,|n\rangle$ is $\Sigma_{i,j} r_ir_j^{\dagger} 
\rho_{ij}$. Thus the ``degenerating locus'' is actually as follows.\\

$$
\begin{array}{ccccc}
V_{A}^k(\rho)=\{(r_1,...,r_m)\in CP^{m-1}:rank( \Sigma_{i,j}r_ir_j^{\dagger} \rho_{ij}) \leq k\}
\end{array}
$$
for $k=0,1,...,n-1$. Similarly $V_{B}^k (\rho) \subseteq CP^{n-1}$ can be defined. It is known from Theorem 1 and 2 of 
[1] that these sets are algebraic sets (zero locus of several multi-variable polynomials, see [21]) and they are 
invariants under local unitary operations depending only on the eigenvectors of $\rho$. Actually these algebraic sets 
can be computed easily as follows.\\

Let $\{|11\rangle,...,|1n\rangle,...,|m1\rangle,...,|mn\rangle \}$ be the standard orthogonal basis of $H_A^m \otimes 
H_B^n$ as above and $\rho= \Sigma_{l=1}^{t} p_l |v_l \rangle \langle v_l|$ be any given  representation of $\rho$ as a 
convex combination of projections with $p_1,...,p_t >0$ (for example, we can take the spectral decomposition 
$\rho=\Sigma_{i=1}^r \lambda_i |\psi_i\rangle \langle \psi_i|$ as a such representation). Suppose 
$v_l=\Sigma_{i,j=1}^{m,n} a_{ijl} |ij\rangle $ , $X=(a_{ijl})_{1\leq i \leq m, 1 \leq j \leq n, 1 \leq l \leq t}$ is 
the $mn \times t$ matrix. Then it is clear that the matrix representation of $\rho$ with the basis 
$\{|11\rangle,...,|1n\rangle,...,|m1\rangle,...,|mn\rangle \}$ is $XPX^{\dagger}$, where $P$ is the diagonal matrix 
with diagonal entries $p_1,...,p_t$. We may consider the $mn\times t$ matrix $X$ as a $m\times 1$ blocked matrix with 
each block $X_w$, where $w=1,...,m$, a $n\times t$ matrix corresponding to $\{|w1\rangle ,...,|wn \rangle\}$. It is 
clear $\rho_{ij}=X_i P X_j^{\dagger}$ and $\Sigma_{i,j} r_ir_j^{\dagger} \rho_{ij}=(\Sigma_i r_i X_i)P(\Sigma_i r_i 
X_i)^{\dagger}$. From simple linear algebra $V_A^k(\rho)$ is just the set of points $(r_1,...,r_m)$ in $CP^{m-1}$ such 
that $rank(\Sigma_i r_iX_i)$ is less than $k+1$, ie., $V_A^k(\rho)$ is the algebraic set in $CP^{m-1}$ as the zero 
locus of the determinants of all $(k+1) \times (k+1)$ submatrices of $\Sigma_i r_i X_i$ (see [1]).\\

We can see that $V_A^k(\rho)$ is independent of eigenvalues $\lambda_i$'s if it is computed from the spectral 
decomposition $\rho=\Sigma_{i=1}^r \lambda_i |\psi_i\rangle \langle\psi_i|$, since it is computed from the matrix $X$ 
depending only on eigenvectors $|\psi_i \rangle$'s.  For example, let $T$ be a two-qubit mixed state with some of the 
following 4 Bell states as its eigenvectors.\\

$$
\begin{array}{ccccccccc}
|v_1 \rangle=\frac{1}{\sqrt{2}}(|11\rangle+|22\rangle)\\
|v_2 \rangle=\frac{1}{\sqrt{2}}(|11\rangle-|22\rangle)\\
|v_3 \rangle=\frac{1}{\sqrt{2}}(|12\rangle+|21\rangle)\\
|v_4 \rangle=\frac{1}{\sqrt{2}}(|12\rangle-|21\rangle)
\end{array}
$$

It is easy to calculate the matrix $r_1 X_1+r_2X_2$ of the whole 4 Bell states, it is the following $2 \times 4$ 
matrix.\\

$$
\left(
\begin{array}{ccccccccc}
r_1&r_1&r_2&-r_2\\
r_2&-r_2&r_1&r_1
\end{array}
\right)
$$

Therefore $V_A^0(T)$ and $V_A^1(T)$ can be computed from the submatrix consisting of $rank(T)$ columns of the above 
matrix. Thus $V_A^0(T)$ is always empty, $V_A^1(T)$ is the set of 2 points when $rank(T)=2$ and empty when 
$rank(T)=3,4$.\\

From [1], Schmidt ranks of pure states $\rho$(ie, projection operator to a unit vector)  are  just the codimensions of 
the algebraic sets ($codim V_A^0(\rho)=codim V_B^0(\rho)$). Therefore it is natural to think the above observation can 
be extended to the equalities of these algebraic sets of arbitrary bipartite semi-positive Hamiltonians of the same 
rank if they can be simulated efficiently. In this paper we give such a  necessary condition about the efficient 
simulation of semi-positive Hamiltonians. \\

{\bf  Theorem 1.} {\em Let $H$ and $H'$ be the semi-positive Hamiltonians on the bipartite quantum system $H_A^m 
\otimes H_B^n$ with the same rank, ie., $dim (range (H))=dim (range (H'))$. Suppose that $H' \prec_{LU} H$, that is , 
$H'$ can be simulated by $H$ efficiently by using local unitary operations. Then $V_A^k(H)=V_A^k(H')$ for 
$k=0,...,n-1$ and $V_B^k(H)=V_B^k(H')$ for $k=0,...,m-1$, here the equality of algebraic sets means they are 
isomorphic via projective linear transformations of complex projective spaces.}\\

The following observation is the the key point of the proof of  Theorem 1. From Lemma 1 in [20] as cited above, the 
range of $\rho$ is the linear span of vectors $|v_1\rangle,...,|v_t\rangle$. We take any $dim(range(\rho))$ linear 
independent vectors in the set $\{|v_1 \rangle,...,|v_t\rangle\}$, say they are $|v_1\rangle,...,|v_s\rangle$ , where 
$s=dim(range(\rho))$. Let $X'$ be the $mn \times s$ matrix with columns corresponding to the $s$ vectors 
$|v_1\rangle,...,|v_s\rangle$'s coordinates in the standard basis of $H_A^m \otimes H_B^n$. Then $X'$ is a submatrix 
of the above-described matrix $X$ and each column of $X$ is a linear combination of columns in $X'$. We consider $X'$ 
as $m \times 1$ blocked matrix with blocks $X'_1,...,X'_m$ ($n \times s$ matrices) as above. It is clear that 
$V_A^k(\rho)$ is just the zero locus of determinants of all $(k+1) \times (k+1)$ submatrices of $\Sigma_i r_iX'_i$, 
since any column in $\Sigma_i r_i X_i$ is a linear combination of columns in $\Sigma_i r_i X'_i$ ( thus $rank(\Sigma_i 
r_i X_i) \leq k$ is equivalent to $rank(\Sigma_i r_iX'_i) \leq k$).\\

{\bf Proof of Theorem 1.} Suppose $H' \prec_{LU} H$, then there exist positive numbers $p_1,...,p_s$ and local unitary 
operations $U_1 \otimes V_1,...,U_t \otimes V_t$, such that, $\Sigma_{i=1}^t p_i U_i \otimes V_i H (U_i \otimes 
V_i)^{\dagger}=H'$. Let $H=\Sigma_{i=1}^s q_i |\psi_i \rangle \langle \psi_i|$, where $s=dim(range(H))$ , 
$q_1,...,q_s$ are eigenvalues of $H$ and $|\psi_1 \rangle ,...,|\psi_s \rangle $ are eigenvectors of $H$. Then it is 
clear that $(U_i \otimes V_i)H(U_i \otimes V_i)^{\dagger}=\Sigma_{j=1}^s q_j |(U_i \otimes V_i)\psi_j \rangle \langle 
(U_i \otimes V_i)\psi_j|$ and thus $H'=\Sigma_{i=1,j=1}^{t,s} p_i q_j |(U_i \otimes V_i)\psi_j \rangle \langle (U_i 
\otimes V_i)\psi_j|$. This is a representation of $H'$ as a convex combination of projections. From Lemma 1 
$range(H')$ is the linear span of $|(U_1 \otimes V_1)\psi_1 \rangle ,...,|(U_1 \otimes V_1)\psi_s \rangle$ since they 
are $dim (range(H'))=s$ linear independent vectors in $range(H')$. From our above observation $V_A^k(H')$ can be 
computed from the matrix $X'$ of vectors $|(U_1 \otimes V_1)\psi_1 \rangle,...,|(U_1 \otimes V_1)\psi_s \rangle$ and 
thus $V_A^k(H')=V_A^k((U_1 \otimes V_1)H(U_1\otimes V_1)^{\dagger})$ from the definition. Thus the conclusion follows 
from Theorem 1 in [1].\\

Since the algebraic-geometric invariants are independent of eigenvalues, thus our above theorem is a necessary 
condition of simulation of Hamiltonians without referring to eigenvalues.  As described in e.g. [11] and [16], local 
terms like $I \otimes K_B$ and $K_A \otimes I$ are considered irrelevant to the simulation process upto the second 
order, this leads to the so-called normal forms of two-qubit Hamiltonians. We can recall the Theorem in section F of 
[11], for Hamiltonians $H$ and $H'$ in their normal forms, ie., $H=\Sigma_i h_i \sigma_i \otimes \sigma_i$ and $H'= 
\Sigma_i h'_i \sigma_i \otimes \sigma_i$, where $\sigma_i$'s are Pauli matrices, on two-qubit systems, $H' \prec_{LU} 
H$ if and only if ${\bf h'}=(h'_1,h'_2,h'_3) \prec_{s} {\bf h}=(h_1,h_2,h_3)$ , where $\prec_{s}$ is the 
s-majorization defined in [11]. Thus we can see that in the case of efficient simulation of Hamiltonians on two-qubit 
systems, eigenvalues of Hamiltonians play a crucial role, since $h$ and $h'$ can be determined from the eigenvalues of 
Hamiltonians $H$ and $H'$ uniquely. However Theorem 1  implies that in the case of arbitrary bipartite quantum 
systems, the algebraic-geometric invariants which are independent of eigenvalues play a more fundamental role. This is 
also  illustrated in the following example of efficient simulation of Hamiltonians in $H_A^3 \otimes H_B^3$.\\

{\bf Example 1.}  Let $|v_1 \rangle,|v_2 \rangle |v_3 \rangle$ be the following 3 unit vectors in $H_{A}^3 \otimes 
H_{B}^3$.\\

$$
\begin{array}{cccccccc}
|v_1 \rangle=\frac{1}{\sqrt{3}}(e^{i\eta_1}|11 \rangle +|22\rangle +|33\rangle )\\
|v_2 \rangle=\frac{1}{\sqrt{3}}(e^{i\eta_2}|12 \rangle +|23\rangle +|31\rangle)\\
|v_3 \rangle=\frac{1}{\sqrt{3}}(e^{i\eta_3}|13\rangle+|21\rangle+|32\rangle)
\end{array}
(1)
$$

,where $\eta_1,\eta_2,\eta_3$ are $3$ real parameters. Let $H_{\eta_1,\eta_2,\eta_3}=(|v_1\rangle\langle v_1| + |v_2 
\rangle \langle v_2|+|v_3\rangle \langle v_3|$. This is a continuous family of Hamiltonians in 
$H_{\eta_1,\eta_2,\eta_3}$ of rank 3 parameterized by three real parameters. \\

It is easy to calculate that  $V_A^2(H_{\eta_1,\eta_2,\eta_3})$ is just the elliptic curve (see [21],[22]) in $CP^2$ 
defined by $r_1^3 +r_2^3 +r_3^3 
-\frac{e^{i\eta_1}+e^{i\eta_2}+e^{i\eta_3}}{e^{i(\eta_1+\eta_2+\eta_3)/3}}r_1r_2r_3=0$. Set 
$g(\eta_1,\eta_2,\eta_3)=\frac{e^{i\eta_1}+e^{i\eta_2}+e^{i\eta_3}}{e^{i(\eta_1+\eta_2+\eta_3)/3}}$ and 
$k(x)=\frac{x^3(x^3+216)^3}{(-x^3+27)^3}$, then $k(g(\eta_1,\eta_2,\eta_3))$ is the moduli function of elliptic 
curves. From algebraic-geometry it is known that if $k(g(\eta_1,\eta_2,\eta_3)) \neq 0,27,-216$, then 
$V_A^2(H_{\eta_1,\eta_2,\eta_3})$ is not the union of 3 lines and when $k(g(\eta_1,\eta_2,\eta_3))=0$ or $27$ or 
$-216$, $V_A^2(H_{\eta_1,\eta_2,\eta_3})$ is the union of 3 lines. Moreover $V_A^2(H_{\eta_1,\eta_2,\eta_3})$ is 
isomorphic to $V_A^2(H_{\eta'_1,\eta'_2,\eta'_3})$ by projective linear transformations if and only if 
$k(g(\eta_1,\eta_2,\eta_3))=k(g(\eta'_1,\eta'_2,\eta'_3))$ ( see section 7.2 , pp.363-396 of [22]). Thus we 
immediately know that $H_{0,0,0}$ cannot be efficiently simulated by $H_{0,0,\pi}$ using local unitary operations from 
Theorem 1. Generally we have the following result.\\

{\bf Corollary 1.} {\em $H_{\eta'_1,\eta'_2,\eta'_3}$ cannot be simulated by $H_{\eta_1,\eta_2,\eta_3}$ efficiently by 
using local unitary transformations,ie.,we cannot have  $H_{\eta'_1,\eta'_2,\eta'_3} 
\prec_{LU}H_{\eta_1,\eta_2,\eta_3}$, if $k(g(\eta_1,\eta_2,\eta_3)) \neq k(g(\eta'_1,\eta'_2,\eta'_3))$, though the 3 
nonzero eigenvalues of $H_{\eta_1,\eta_2,\eta_3}$,\\$H_{\eta'_1,\eta'_2,\eta'_3}$ and their partial traces are all 
1.}\\

{\bf Proof.} It is easy to calculate the eigenvalues to check the 2nd conclusion. The first conclusion is from Theorem 
1 and the above-described well-known fact about elliptic curves.\\

This example strongly suggests that the problem of efficient simulation of Hamiltonians  on arbitrary bipartite 
quantum systems is quite different to the problem in two-qubit case as studied in [11].\\

Let $S$ be the swap operator on the bipartite system $H_A^n \otimes H_B^n$ defined by $S|ij\rangle=|ji\rangle$. For 
any Hamiltonian $H$, $S(H)=SHS^{\dagger}$ corresponds to the Hamiltonian evolution of $H$ with A and B interchanged. 
It is very interesting to consider the problem if $H$ can be simulated by $S(H)$ efficiently . This led to some 
important consequences in the discussion VII of [11]. For example it was shown there are examples that $H$ and $S(H)$ 
cannot be simulated efficiently with one another in higher dimensions. Thus in higher dimensions nonlocal degrees of 
freedom of Hamiltonians cannot be characterized by quantities that are symmetric with respect to A and B, such as 
eigenvalues. This conclusion  is also obtained from the above Corollary 1. From Theorem 1 we have the following 
necessary condition about $H \prec_{LU} S(H)$.\\

{\bf Corollary 2.} {\em Let $H$ be a semi-positive Hamiltonian on $H_A^n \otimes H_B^n$. Suppose $H \prec_{LU} S(H)$. 
Then $V_A^k(H)=V_B^k(H)$ for $k=0,...,n-1$.}\\

The following is a Hamiltonian $H$  on $3 \times 3$ system for which $H$ cannot be simulated efficiently by $S(H)$.\\

{\bf Example 2.} $H=|\phi_1\rangle \langle\phi_1|+|\phi_2\rangle \langle \phi_2|+|\phi_3\rangle \langle\phi_3|$, 
where,\\

$$
\begin{array}{cccccc}

|\phi_1 \rangle=\frac{1}{\sqrt{3}}(|11\rangle +|21\rangle +|32\rangle)\\
|\phi_2 \rangle=\frac{1}{\sqrt{1+|v|^2}}(|12\rangle+v|22\rangle)\\
|\phi_3 \rangle=\frac{1}{\sqrt{1+|\lambda|^2}}(|13\rangle +\lambda|23\rangle)\\
\end{array}
(2)
$$

Then it is easy to compute that $V_A^2(H)$ is the sum of 3 lines in $CP^2$ defined by $r_1+r_2=0$,$r_1+vr_2=0$ and 
$r_1+\lambda r_2=0$ for $v \neq \lambda$ and both $v ,\lambda$ are not 1, and $V_B^2(H)$ is the sum of 2 lines in 
$CP^2$ defined by $r_2=0$ and $r_3=0$. Thus we cannot have $H \prec_{LU} S(H)$.\\

The following example shows that our results can lead to non-trivial constraints without referring to eigenvalues even 
in the two-qubit case if local terms are not neglected (as in our definition). Although local terms may be physically 
irrelevant in the setting of [11] and [16], Example 3 below illustrates mathematically how our results work.\\

{\bf Example 3.} Let $H=\lambda_1|\psi_1 \rangle \langle \psi_1|+\lambda_2|\psi_2 \rangle \langle \psi_2|$ and  
$H'=\lambda'_1|\psi'_1 \rangle \langle \psi'_1|+\lambda'_2|\psi'_2 \rangle \langle \psi'_2|$ be two Hamiltonians on 
$H_A^2 \otimes H_B^2$, where $\lambda$'s are any given positive real numbers such that $trH=trH'$ and \\

$$
\begin{array}{cccccccc}
|\psi_1 \rangle=\frac{1}{\sqrt{2}}(|11\rangle+|22\rangle)\\
|\psi_2 \rangle=\frac{1}{\sqrt{2}}(|11\rangle-|22\rangle)\\
|\psi'_1 \rangle=\frac{1}{\sqrt{2}}(|11\rangle+|22\rangle)\\
|\psi'_2\rangle=|12 \rangle\\
\end{array}
(3)
$$

Then we know $H$ and $H'$ are two rank 2 Hamiltonians. It is easy to compute that $V_A^1(H)$ is the algebraic set of 
two points $(1:0)$ and $(0:1)$ in $CP^1$ and $V_A^1(H')$ is the algebraic set of one point $(0:1)$ in $CP^1$. Hence we 
cannot have $H' \prec_{LU} H$ from the Theorem 1.\\

We can now observe the compatibility of our necessary condition Theorem 1 with the sufficient and necessary condition 
in two-qubit case in [11]. For two two-qubit Hamiltonians $H$ and $H'$ in their normal forms, i.e., $H=\Sigma_i h_i 
\sigma_i \otimes \sigma_i$ and $H'=\Sigma_i h'_i \sigma_i \otimes \sigma_i$, it is proved in [11] that $H' \prec_{LU} 
H$ if and only if ${\bf h'}=(h'_1,h'_2,h'_3) \prec_s {\bf h}=(h_1,h_2,h_3)$, i.e., ${\bf h'}$ is s-majorized by ${\bf 
h}$.  It is clear that the 4 eigenvectors of any two-qubit Hamiltonian in its normal form are exactly 4 Bell states. 
Thus if $H=\Sigma_i h_i \sigma_i \otimes \sigma_i$ is of the form $T-\frac{trT}{4}I_4$, where $T$ is a semi-positive 
two-qubit Hamiltonian, then the algebraic-geometric invariants of $T$ are fixed, i.e., $V_A^0(T)$ is empty, and 
$V_A^1(T)$ is the set of 2 points when $rank (T)=2$ and empty when $rank (T)=3$ or 4. Thus we can see that our 
necessary condition Theorem 1 is void when applied to two-qubit Hamiltonians in their normal forms, the necessary 
condition in this paper is compatible with the main result in [11].\\

Actually the algebraic geometric invariants in [1] can be used to give more necessary conditions for the efficient 
simulation of Hamiltonians by using local unitary operations.\\

{\bf Theorem 2.} {\em Let $H$ and $H'$ be two semi-positive Hamitonians on $H_A^m \otimes H_B^n$. Suppose that there 
exists a representation of $H$ as a convex combination $H=\Sigma_i^s q_i |v_i\rangle \langle v_i|$, with positive 
$q_i$'s and the Schmidt rank of $|v_1\rangle$ is $min\{m,n\}$. Moreover $V_A^0(H')$ is not empty. Then $H'$ cannot be 
simulated by $H$ efficiently by using local unitary operations, ie., we cannot have $H' \prec_{LU} H$.}\\

{\bf Proof.} From the condition, there exist positive $p_1,...,p_t$ and local unitary operations $U_1 \otimes 
V_1,...,U_t \otimes V_t$, such that, $\Sigma_{i=1}^t p_i U_i \otimes V_i H (U_i \otimes V_i)^{\dagger}=H'$. It is 
clear that $(U_i \otimes V_i)H(U_i\otimes V_i)^{\dagger}=\Sigma_{j=1}^s q_j |(U_i \otimes V_i)v_j \rangle \langle(U_i 
\otimes V_i)v_j|$, and thus $H'=\Sigma_{i=1,j=1}^{t,s} p_iq_j |(U_i \otimes V_i)v_j \rangle \langle (U_i \otimes 
V_i)v_j|$. From Lemma 1 in [20] as cited ( Lemma 1 ), $range (H')$ is the linear span of vectors $(U_i \otimes 
V_i)v_j$ for $i=1,...,t$ and $j=1,...,s$. From the above description about the computation of $V_A^0(H')$, we can 
compute it by choosing $dim(range(H'))$ linear independent vectors in this set $\{(U_1 \otimes V_1)v_1,...,(U_1 
\otimes V_1)v_s,...,(U_t \otimes V_t)v_1,...,(U_t \otimes V_t) v_s\}$. Therefore we can choose one of these 
$dim(range(H'))$ linear independent vectors to be $(U_1 \otimes V_1)v_1$, whose Schmidt rank is $min\{m,n\}$.  From 
[1] and the definition , we know that $V_A^0(H')$ has to be the empty set. This is a contradiction and the conclusion 
is proved.\\

{\bf Example 4.} Let $H=|v \rangle \langle v|$ and $H'=\frac{1}{2}(|u_1 \rangle \langle u_1|+|u_2 \rangle \langle 
u_2|)$ be two Hamiltonians on $H_A^3 \otimes H_B^3$ where\\

$$
\begin{array}{cccccccccc}
|v \rangle=\frac{1}{\sqrt{3}}(|11\rangle +|22\rangle +|33 \rangle)\\
|u_1\rangle=\frac{1}{\sqrt{2}}(|11\rangle +|22\rangle)\\ 
|u_2\rangle=\frac{1}{\sqrt{2}}(|11\rangle -|22\rangle )\\ 
\end{array}
(4)
$$

It is clear that $V_A^0(H')$ is the set of one point $(0:0:1)$ in $CP^3$, thus nonempty. On the other hand $H$ 
satisfies the condition in Theorem 2. Thus we cannot have $H' \prec_{LU} H$.\\

For multipartite Hamiltonians on $H_{A_1}^{m_1} \otimes \cdots \otimes H_{A_n}^{m_n}$, the definition $ H' \prec_{LU} 
H$ can be naturally extended as follows. We say that multipartite Hamiltonian $H'$ can be simulated by $H$ efficiently 
using local unitary operations, written as $H' \prec_{LU} H$, if there are positive real numbers $p_1,...,p_s$ such 
that $p_1+..+p_s=1$ and unitary operations $U_1^1,...,U_s^1$,...,$U_1^n,...,U_s^n$ on 
$H_{A_1}^{m_1}$,...,$H_{A_n}^{m_n}$ respectively, such that, $H'=p_1(U_1^1 \otimes \cdots \otimes U_1^n)H(U_1^1 
\otimes \cdots \otimes U_1^n)^{\dagger}+...+p_s(U_s^1 \otimes \cdots \otimes U_s^n)H(U_s^1 \otimes \cdots \otimes 
U_s^n)^{\dagger}$. Then we can use the algebraic-geometric invariants in [2] to give the following necessary 
condition.\\

{\bf  Theorem 3.} {\em Let $H$ and $H'$ be the semi-positive Hamiltonians on the multipartite quantum system 
$H_{A_1}^{m_1} \otimes \cdots \otimes H_{A_n}^{m_n}$ with the same rank. Suppose that $H' \prec_{LU} H$, that is , 
$H'$ can be simulated by $H$ efficiently by using local unitary operations. Then $V_{A_{i_1}:...:A_{i_j}}^k(H)=V_{ 
A_{i_1}:...:A_{i_j}}^k(H')$ for any possible $k$ and any possible $A_{i_1},...,A_{i_j}$ , here the equality of 
algebraic sets means they are isomorphic via projective linear transformations of the product of 
 complex projective spaces.}\\

The proof is the same as  the proof of Theorem 1.\\

The following example is 3 qubit case.\\

{\bf Example 5.} Let $H$ and $H'$ be rank 4 Hamiltonians on $H_A^2 \otimes H_B^2 \otimes H_C^2$, $H=|\phi_1 \rangle 
\langle \phi_1|+|\phi_2 \rangle \langle \phi_2|+|\phi_3 \rangle \langle \phi_3|+|\phi_4 \rangle \langle \phi_4|$ and 
$H'=|\phi'_1 \rangle \langle \phi'_1|+|\phi'_2 \rangle \langle \phi'_2|+|\phi'_3 \rangle \langle \phi'_3|+|\phi'_4 
\rangle \langle \phi'_4|$, where, \\

$$
\begin{array}{cccccccc}
|\phi_1 \rangle=\frac{1}{\sqrt{2}}(|010\rangle -|011\rangle)\\
|\phi_2 \rangle=\frac{1}{\sqrt{2}}(|100\rangle -|110\rangle)\\
|\phi_3 \rangle=\frac{1}{\sqrt{2}}(|001\rangle -|101\rangle)\\
|\phi_4 \rangle=\frac{1}{\sqrt{2}}(|000\rangle-|111\rangle)\\

|\phi'_1 \rangle=\frac{1}{\sqrt{2}}(|000\rangle-|100\rangle)\\
|\phi'_2 \rangle=\frac{1}{\sqrt{2}}(|001\rangle-|101\rangle)\\
|\phi'_3 \rangle=\frac{1}{\sqrt{2}}(|010\rangle-|110\rangle)\\
|\phi'_4 \rangle=\frac{1}{\sqrt{2}}(|011\rangle-|111\rangle)\\
\end{array}
(5)
$$

Then we can compute that $V_{A:B}^1(H)$ is the sum of $CP^1 \times (1:0),(0:1)\times CP^1$ and $(1:0) \times (0:1)$ in 
$CP^1 \times CP^1$, and $V_{A:B}^1(H')$ is the set of two points $(1:1) \times (0:1)$ and $(1:1) \times (0:1)$ in 
$CP^1 \times CP^1$. Thus from Theorem 3, we cannot have $H' \prec_{LU} H$.\\

In conclusion, we have proved necessary conditions for the efficient simulation of both bipartite and multipartite 
Hamiltonians using local unitary operations, which are independent of eigenvalues and based on algebraic-geometric 
invariants. These conditions indicated that the in higher dimension bipartite cases or multipartite cases, the 
relation of efficient simulation of Hamiltonians depends more on the eigenvectors than eigenvalues. This is quite 
different to the two-qubit case studied in [11]. \\

It is natural to ask if the techniques from algebraic geometry used here can be extended to find not only necessary 
but also sufficient conditions of simulating semi-positive bipartite Hamiltonians by another with local unitary 
operations. For low rank semi-positive bipartite Hamiltonians or semi-positive Hamiltonians on low dimensional 
bipartite systems, it seems that the eigenvalues of these Hamiltonians, eigenvalues of their partial traces and 
algebraic-geometric invariants are near a complete set of invariants under local unitary operations, ie., we almost 
can determine exactly in which equivalent class the bipartite Hamiltonians are if we know all these invariants. Thus 
in these cases it seems hopeful to extend the techniques here to find necessary and sufficient conditions of 
simulation problem of Hamiltonians. However in general case we think it would be difficult to get necessary and 
sufficient conditions about this problem based on present-known invariants. We speculate that more invariants of 
bipartite Hamiltonians under local unitary operations have to be found for the purpose to completely describe the 
ability of bipartite Hamiltonians to simulate one another with the help of local unitary operations. \\

{\bf Acknowledgment}: The author thanks two referees for their careful reading of the manuscript and very helpful 
comments and criticisms. The author acknowledges the support from NNSF China, Information Science Division, grant 
69972049 and the "Distinguished Young Scholar Grant" 10225106 .\\

e-mail: chenhao1964cn@yahoo.com.cn or mcsch@zsu.edu.cn\\

\begin{center}
REFERENCES
\end{center}

1.H. Chen, quant-ph/0108093\\

2.H. Chen, quant-ph/0109056\\

3.R.P.Feynman, Int. J. Ther. Phys., 21:467, 1982\\

4.H.Rabitz, R.de Vivie-Riedle, M.Motzkus and K.Kompa, Science 288, 824 (2000)\\

5.N.Linden, H.Barjat, R.Carbajo, and R.Freeman, Chemical Physics Letters, 305:28-34, 1999\\

6.D.W.Leung, I.L.Chuang, F.Yamaguchi and Y.Yamamoto, Phys. Rev. A, 61:042310, 2000\\

7.J.Jones and E.Knill, J. Mag. Res., 141:322-5, 1999\\

8.J.L.Dodd, M.A.Nielsen, M.J.Bremner and R.T. Thew, quant-ph/0106064\\

9.W.D\"{u}r, G.Vidal, J.I.Cirac, N.Linden and S. Popescu, Phys. Rev. Lett., 87:137901 (2001)\\

10.P.Zanardi, C.Zalka and L.Faoro, quant-ph/0005031\\

11.C.Bennett, J.I.Cirac, M.S.Leifer, D.W.Leung, N.Linden, S.Popescu and G.Vidal, quant-ph/0107035,v2, 
Phys.Rev.A,66,012305 (2002)\\

12.D.W.Leung, quant-ph/0107041, to appear in J.Mod.Opt.\\

13.D.Janzing,quant-ph/0108052\\

14.D.Janzing and Th.Beth, quant-ph/0108053\\

15.M.A.Nielsen,M.J.Bremner,J.L.Dodd,A.M.Childs,C.M.Dawson, quant-ph/0109064\\

16.P.Wocjan, M.Roetteler, D.Janzing and Th.Beth, quant-ph/0109063, Quantum Information and Computation, 
Vol.2(2002),No.5,133-150,
 quant-ph/0109088, Phys.Rev. A, 65, 042309(2001)\\

17.G.Vidal and J.I.Cirac, quant-ph/0108076\\

18.M.A.Nielsen,Phys.Rev.Lett.,83(1999) 436\\

19.N.Linden,S.Popescu and A.Sudbery, Phys. Rev. Lett, 83(1999), 243\\

20.P.Horodecki, Phys.Lett.A 232(1997)\\

21.J.Harris,Algebraic geometry, GTM 133, Springer-Verlag, 1992\\

22. E.Brieskorn and H.Knorrer, Ebene algebraische Kurven, Birkh\"{a}user, Basel-Boston-Stuttgart, 1981\\

\end{document}